\documentclass[12pt]{article}
\usepackage{a4,amsmath,amssymb,cite,graphicx}
\usepackage{rotating}
\usepackage{color}
\newcommand{\node}[4]{\begin{smallmatrix} #1\\ #2\\  #3 \\ #4 \end{smallmatrix}}
\catcode`@=11 \@addtoreset{equation}{section} \catcode`@=12

\begin{document}
\bibliographystyle{utphys}
\begin{titlepage}
\renewcommand{\thefootnote}{\fnsymbol{footnote}}
\noindent
\phantom{\tt IITM/PH/TH/2013/8}\hfill
{\tt arXiv:AAAA.BBBB} \\[4pt]
\mbox{}\hfill 
\hfill{\fbox{\textbf{v3.01; June 2014 }}}

\begin{center}
\large{\sf  Estimating the asymptotics of solid partitions}
\end{center} 
\bigskip 
\begin{center}
{\sf Nicolas Destainville\footnote{destain at irsamc.ups-tlse.fr} } \\[3pt]
\textit{Universit\'e de Toulouse, UPS\\
CNRS; LPT (IRSAMC)\\
 Laboratoire de Physique Th\'eorique\\
 F-31062 Toulouse, France, EU\\[4pt]}
\end{center}
\centerline{and}
\begin{center}
{\sf Suresh Govindarajan\footnote{suresh at physics.iitm.ac.in} } \\[3pt]
\textit{Department of Physics,\\ Indian Institute of Technology Madras,\\ Chennai 600036, India \\[4pt]
}
\end{center}
\bigskip
\bigskip
\begin{abstract}
We study the asymptotic behavior of solid partitions using transition matrix Monte Carlo simulations. If $p_3(n)$ denotes the number of solid partitions of an integer $n$, we show that $\lim_{n\rightarrow\infty} n^{-3/4} \log p_3(n)\sim 1.822\pm 0.001$. This shows clear deviation from the value $1.7898$, attained by MacMahon numbers $m_3(n)$,  that was conjectured to hold for solid partitions as well. In addition, we find estimates for other sub-leading terms in $\log p_3(n)$. In a pattern deviating from the asymptotics of line and plane partitions, we need to add an oscillatory term in addition to the obvious  sub-leading terms. The period of the oscillatory term is proportional to $n^{1/4}$, the natural scale in the problem. This new oscillatory term might shed some insight into why partitions in dimensions greater than two do not admit a simple generating function.
\end{abstract}
\end{titlepage}
\setcounter{footnote}{0}

\section{Introduction}

The partitions of integers and their higher dimensional generalizations are simple to define and yet provide a rich structure that continues to fascinate scientists by appearing in diverse fields. As the dimensionality of the partitions increases, our understanding progressively gets reduced. The usual (one-dimensional) partitions are the best understood with a generating function due to Euler that has connections with modular forms. There is an elegant exact formula due to Hardy-Ramanujan-Rademacher (HRR) that has its origins in an asymptotic formula for partitions of large integers\cite{Andrews1998book}. The generating function of plane (two-dimensional) partitions, while known, is not a modular form and one can obtain its asymptotic behavior, for instance, using the Meinardus method\cite{Mutafchiev06,Wright1931}. There is an attempt at a HRR-type formula due to Almkvist which has been extended recently\cite{alm1,alm2,Govindarajan2013}. For dimensions greater than two, starting with solid (three-dimensional) partitions, no formulae for the generating function are known  and we know the number of such partitions for small values of integers. The largest known is the number of solid partitions of $72$ whose enumeration needed about half a million CPU hours\cite{solidpartitionsproject}. This paper addresses another aspect of solid partitions by using Monte Carlo methods to obtain the asymptotic behavior of solid partitions.

Studying partitions of integers has motivations in various scientific fields, notably in physics (e.g., the $q \rightarrow \infty$ Potts model, directed compact lattice animals, crystal growth, Bose-Einstein statistics, dimer coverings (a.k.a. perfect matchings), as detailed in Reference~\cite{Mustonen03} and references therein). We discuss two other applications that are of interest to us and have lead to this collaborative effort.

Even though plane or solid partitions where studied by mathematicians since the early twentieth century \cite{MacMahon16}, the discovery of quasicrystals in 1984 prompted renewed interest in these fascinating combinatorial objects, because of the connection between partitions and Penrose-like random tilings, which have themselves rapidly been identified as simplified atomistic models of quasicrystals~\cite{Levine84,Elser85}. Enumerating plane, solid, or even generalized partitions\cite{Mosseri93,octo01,Widom02,Destainville04,largecodim05,Widom13} has become a popular objective because the resulting extensive configurational entropy has been identified as a serious candidate to account for quasicrystal thermodynamic stability against competing crystal phases. Stability might also arise from quasicrystal electronic properties, which also motivated the study of quantum transport in random tilings (see Reference~\cite{Vidal03} and references therein).

The generating function of $d$-dimensional partitions naturally appears in the counting of BPS states (and supersymmetric black holes) for type IIA string theory with target space, $\mathbb{C}^{d+1}$, as shown by Gopakumar and Vafa\cite{Gopakumar98a}.  These turn out to be related to the Hilbert scheme of points on $\mathbb{C}^{d+1}$ with the corresponding partitions giving the Euler characteristic of the Hilbert scheme. The generating functions for more general target spaces, in dimensions two and three, appear as deformations of the generating function by additional parameters corresponding to the K\"ahler moduli of these spaces\cite{Gopakumar98a,Gopakumar98b,Behrend2009}. For target spaces with dimension four such as $\mathbb{C}^4$, where the possibility of solid partitions and its generating function appearing, the answers are however not  well-understood and are of current research interest.

The lack of any systematic understanding of solid partitions and their higher dimensional counterparts has been a hurdle in studies of problems where they appear. A ray of hope appeared in the form of a conjectures of \cite{Mustonen03,Balakrishnan2011} on the asymptotics of solid and higher-dimensional partitions. The idea was that a formula guessed by MacMahon for the generating function of higher-dimensional partitions  might correctly reproduce the asymptotic behavior  even though it has been proven to be wrong\cite{Atkin63}. The work of \cite{Widom02} suggested that a similar conjecture for the configurational entropy based on a similar formula, again due to MacMahon, was incorrect.  This work aims at using high quality Monte Carlo simulations to establish the asymptotic behavior of solid partitions to test the conjectures of  \cite{Mustonen03,Balakrishnan2011}.

The paper is organized as follows. Following the introductory section, in section 2, we briefly provide the necessary background and definitions for our problem. Section 3 is the main part of the paper, where we discuss the details of the Monte Carlo simulation that we used, the motivation for the formula used in the fit and the results of the fit.  Section 3.5 is devoted to showing the unanticipated  appearance of oscillatory behaviour at sub-sub-leading order that is visible due to the high quality of our data. We conclude in section 5 with a summary and some remarks. Appendix A focuses on the asymptotic of plane partitions and in appendix B, we discuss the asymptotics of MacMahon numbers. Finally, in appendix C, we provide some evidence for oscillatory behaviour for solid partitions that are restricted to a box.

\section{Background}
\label{background}

A $d$-dimensional partition of $n$ is a collection $(X_k)$ of integers~-- where $k$ is a meta-index running over $\mathbb{N}^d$~-- which are weakly decreasing in each direction of space, and such that
\begin{equation} 
\sum_{k \in \mathbb{N}^d} X_k = n.
\label{parts}
\end{equation}
Let $p_d(n)$ denote the number of  $d$-dimensional partitions of $n$.  In this notation, $d=1$ corresponds to the usual partitions, $d=2$ corresponds to plane partitions and $d=3$ corresponds to solid partitions that are the main focus of this paper. There is a second representation of a $d$-dimensional partition as a $(d+1)$-dimensional Ferrers diagram. A Ferrers diagram for a $d$-dimensional partition of $n$ is a collection of $n$ points or \textit{nodes}, $\lambda=(\mathbf{y}_1,\ldots,\mathbf{y}_n)$, in $\mathbb{Z}_{\geq0}^{d+1}$ satisfying the condition\cite{Atkin63}: 
\begin{center}
If  the node $\mathbf{a}=(a_1,a_2,\ldots, a_{d+1})\in \lambda$, then so do all the nodes $\mathbf{y}=(y_1,y_2,\ldots,y_{d+1})$ with $0\leq y_i\leq a_i$ for all $i=1,\ldots, d+1$.
\end{center}
For instance, the Ferrers diagram
\[
\left(\node0000
\node0010
\node0100
\node1000
\node1100
\node2000\right) \ ,
\]
where each column is a node, represents  a solid partition of $6$.

The generating function of $d$-dimensional partitions is defined as follows
\begin{equation}
P_d(q)= 1+ \sum_{n=1}^\infty p_d(n)\ q^n \ .
\end{equation}
MacMahon  conjectured that $P_3(q)$ could be generalized from its plane partition counterpart~\cite{MacMahon16}. This conjecture has later been disproved by exhibiting a counter-example \cite{Atkin63}. 

Even though the generating function for solid partitions is not known, it has been shown that $n^{-(d+1)/d}\log p_d(n)$ has a finite limit~\cite{Bathia97}, that we denote by $\alpha_d$. It has later been proposed that even though incorrect, MacMahon's conjecture might be asymptotically exact, thus providing the Ansatz value  $\alpha_3\simeq 1.78982$ in addition to values for coefficients of the sub-leading terms\cite{Mustonen03,Balakrishnan2011}.  This Ansatz has been supported by numerical investigations~\cite{Mustonen03}. The goal of this work is to better establish the asymptotic behavior of solid partitions in addition to obtaining an improved estimate for $\alpha_3$.

\section{The asymptotics of solid partitions}

\subsection{The solid partition graph}

Consider a  directed graph whose vertices are solid partitions. Two vertices, say $\lambda_1$ and $\lambda_2$,  of the graph are connected by an edge if the Ferrers diagram of the vertex $\lambda_1$  can be obtained from the second vertex, $\lambda_2$ by the addition or deletion of a single node. Direct the edge towards the solid partition with larger number of  nodes. This is the \textit{solid partition graph}. The root vertex of the solid partition graph  is the unique partition with one node and the depth of a given vertex is  the number of nodes in  the partition.  Let Tot$(n)$ denote the total number of incoming edges at depth $n$. For instance, one has Tot$(2)=4$ as all four solid partitions of $2$ are connected to the root vertex. Define Tot$(1)\equiv 1$. Table \ref{Totvalues} lists the values of Tot$(n)$ for $n\leq 40$ -- these were determined numerically using the Bratley-McKay  algorithm to generate all solid partitions of $n$ and then counting the number of incoming edges for the generated solid partition\cite{Bratley:1967a}.

\begin{table}[htbp]
\centering
\begin{tabular}{rr|rr|rr} \hline
$n$ & Tot$(n)$ & $n$ & Tot$(n)$& $n$ & Tot$(n)$\\[3pt] \hline
1& 1 &16& 1100411 & 31& 19625001436\\
2& 4 &17& 2245118 &32& 35765137033 \\
3& 16 &18& 4528212 &33& 64853219808 \\
4& 46 &19& 9038898 &34& 117031972499 \\
5& 128 &20& 17868025 &35& 210211082354 \\
6& 332 &21& 35006932&36& 375886565558 \\
7& 842 &22& 68008606& 37& 669232663688 \\
8& 2042 &23& 131083778 &38& 1186538314110 \\
9& 4846 &24& 250774482&39& 2095236499224\\
10& 11146 &25& 476372848& 40&3685445929502  \\
11& 25114 &26& 898837825&\\
12& 55310 &27& 1685107392& \\
13& 119662 &28& 3139812791&\\
14& 254354 &29& 5816015908&\\
15& 532784&30& 10712596279&   \\ \hline
\end{tabular}
\caption{Exact values of Tot$(n)$ for $n\leq 40$.} \label{Totvalues}
\end{table}

Let $\lambda$ denote a solid partition of $n$ -- we denote this symbolically  by $\lambda\vdash n$.  Let $n_+(\lambda)$ denote the
number of solid partitions  that can reached by adding a node to $\lambda$ (equivalently the number of outgoing edges  at the vertex $\lambda$ in the solid partition graph) and
$n_-(\lambda)$ denote the number of solid partitions that can be reached by deleting a
node from $\lambda$ (equivalently the number of incoming edges  at the vertex $\lambda$ in the solid partition graph). Define $N_\pm(n)$ as follows:
\begin{equation}
\begin{split}
N_+(n):=\frac{\sum_{\lambda\vdash n} n_+(\lambda)}{\sum_{\lambda\vdash n} 1 } =
\frac{\sum_{\lambda\vdash n } n_+(\lambda)}{p_3(n) } \ , \\
N_-(n):=\frac{\sum_{\lambda\vdash n } n_-(\lambda)}{\sum_{\lambda\vdash n} 1 } =
\frac{\sum_{\lambda\vdash n } n_-(\lambda)}{p_3(n) } \ .
\end{split}
\end{equation}
Thus $N_+(n)$ (resp. $N_-(n)$) counts the average number of outgoing (resp. incoming)  edges at depth $n$ in the solid partition graph.
For $n\geq1$, one has the identity
\begin{equation}
 N_-(n)\ p_3(n) =N_+(n-1)\ p_3(n-1) \ . \label{ntop}
\end{equation}
The left hand side of the above equation counts the total number of  incoming edges at depth $n$ i.e, Tot$(n)$ while the right hand side counts the number of outgoing edges at depth $(n-1)$ -- the equality follows since every outgoing edge at depth $(n-1)$ is also an incoming edge at depth $n$. Given $N_{\pm}(n)$, we can
recursively obtain $p_3(n)$ starting from $p_3(1)=1$. One thus obtains, for $n>1$,
\begin{equation}
 p_3(n)=\prod_{m=1}^{n-1} \tfrac{N_+(m)}{N_-(m+1)}\ .
\end{equation}
Suppose we know the number of solid partitions of $n_0$, then for $n>n_0$, we can write
\begin{equation}
 p_3(n)=\left(\prod_{m=n_0}^{n-1} \tfrac{N_+(m)}{N_-(m+1)}\right)\times  p_3(n_0)\ . \label{nnaught}
\end{equation}
For instance, we can choose $n_0$ to be $72$ which is the largest exactly known number of solid partitions~\cite{solidpartitionsproject}.

\subsection{The Monte Carlo Simulation}
\label{MC}

The transition matrix Monte Carlo technique used in this work is adapted from Reference~\cite{Widom02}. Consider solid partitions that are restricted to fit into a cubical box of size $B$ (i.e., $k \in [0,B-1]^3$ and $X_k \leq B$ for all $k$ or, equivalently, all coordinates of all nodes are $\leq B$). This converts the solid partition graph  into one with finite nodes. The solid partition, fitting into the box, and with the largest number of nodes is the unique one with $B^4$ nodes (i.e., $\forall k\in [0,B-1]^3$, $X_k=B$). Define $n_\pm^{\textrm{rest}}(\lambda)$ and $N_\pm^{\textrm{rest}}(n)$ analogous to the unrestricted partitions. 

By tuning an abstract ``temperature'' (see~\cite{Widom02} for further details), the Monte Carlo program traverses the solid partition graph  from depth 1 to some maximum depth, say $n_{max}$ and then back to depth 1 several times. In the process, we obtain estimates 
for  $N_\pm^{\textrm{rest}}(n)$ as follows:
\begin{equation}
 N^{\textrm{rest}}_+(n)\sim \frac{\langle n^{\textrm{rest}}_+\rangle_n}{\textrm{hits}(n)} 
 \quad \textrm{and}\quad
 N^{\textrm{rest}}_-(n)\sim \frac{\langle n^{\textrm{rest}}_-\rangle_n}{\textrm{hits}(n)}  \ ,
\end{equation}
where 
$$
\langle n^{\textrm{rest}}_\pm\rangle_n = \sum_{\lambda\vdash n} n_\pm (\lambda) 
 \quad \textrm{and}\quad
\textrm{hits}(n)= \sum_{\lambda\vdash n}1 \ .
 $$
 with all the above sums running over partitions $\lambda$ visited during the Monte Carlo run. Thus, $\textrm{hits}(n)$ count the total number of solid partitions with $n$ nodes visited during the run.

\begin{figure}[htbp]
\centering
\includegraphics[width=2.5in]{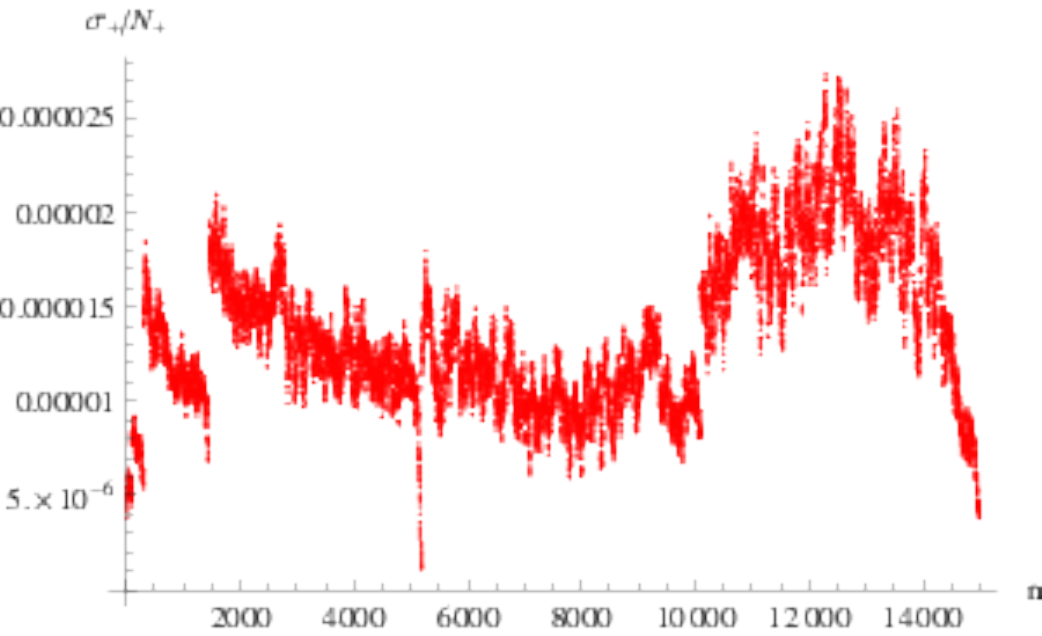}\qquad
\includegraphics[width=2.5in]{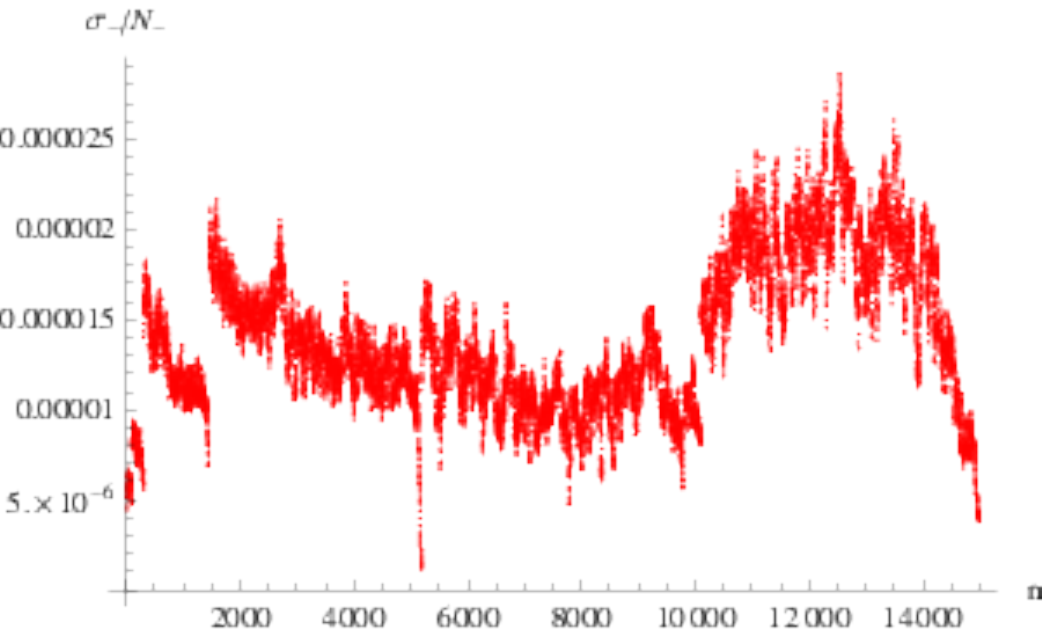}
\caption{Statistical Errors in $N_\pm$ -- this is estimated from the fluctuations in $N_\pm$ using the independent MC runs. It is easy to see the different data sets used. }
\label{errors}
\end{figure}

In order to estimate statistical errors, we carried out several runs with distinct random seeds and different values of $n_{max}$. We carried out 10 runs each for $n_{max}=100,\ 300$ and 25 runs each for $n_{max}=1500,\ 2700,\ 5200,\ 10200,\ 15100$.  The longest runs were for $n_{max}=10200,\ 15100$ -- each run was about 1000 hours. This was necessary for us to reduce the statistical errors.
The data from the $n_{max}=15100$ was \textit{not} used to estimate the asymptotic parameters but as an independent data set to see the goodness of our fits beyond $n=10200$. Let $\sigma_\pm(n)$ denote standard deviation of $N_\pm(n)$ for the $B=300$ dataset. In figure \ref{errors}, we see that $\sigma_\pm/N_\pm$ lies in the range $5\times 10^{-6}$ (for low $n$) to $2.5\times 10^{-5}$ (for large $n$). A similar result is obtained for $B=200$ as well.

We need to understand the effects of finite size on the estimates for the asymptotic behavior. Intuitively, for $n\ll B^4$, we expect $N_\pm(n)\sim N^{\textrm{rest}}_\pm(n)$ since the number of solid partitions that do not fit into the box are exponentially small compared to those that fit into the box.  
In order to quantitatively estimate finite size effects, we have carried out two sets of runs with $B=200$ and $B=300$.  Our fits were carried out  for $n_{max}=10100$, thus $n_{max}/B^4< 10^{-5}$ is reasonably small. We find the finite size effects are comparable to the statistical error in the parameters. This is consistent with our expectation that finite size effects will not be significant due to the small value of $n_{max}/B^4$.

The chosen values of $B$ are also justified by the fact, exemplified later on Figure~\ref{example} (Left), that a typical partition of $n \approx 10^5$ fits into a box of size $B\simeq 100$. For $n\approx10^4$, the typical box size will be even smaller, and partitions not fitting in a box of size $B=200$ or more will be extremely rare.

We also tested the error estimates by  estimating $p_3(n)$ for $n\in[n_0+1,72]$
assuming a value for $n_0$. For instance, with $n_0=50$, using formula Eq \eqref{nnaught} we see that the error is given by adding the statistical errors in $N_\pm(m)$ for all $51\leq m \leq n$ in quadrature. Thus, for $n=72$, the statistical error  is $1.5\times 10^{-3}\%$ while the actual devation from $p_3(72)$ turned out to be $2.3\times 10^{-4}\%$.  We obtain
\begin{align*}
mc_3(72)=\underline{34642}83075820104704\ ,\\
\phantom{m}p_3(72) = 3464274974065172792\ .
\end{align*}
where we have denoted by $mc_3(n)$, the Monte Carlo estimate for $p_3(n)$ and underlined the number of digits that are expected to agree based on our error estimates. The statistical errors in $mc_3(10000)$ and $mc_3(15000)$ are $0.17\%$ and $0.25\%$ respectively. Note that the  error in $mc_3(n)$ increases with $n$ due to the cumulative nature of the statistical errors in $N_\pm(m)$ in Eq. \eqref{nnaught}. This is a drawback if we wish to estimate parameters in an asymptotic formula for $p_3(n)$. However, this is easily addressed as we will see next.

\subsection{Fitting  the data}

The number of solid partitions is a derived quantity not directly computed
by our Monte Carlo program. As the Monte Carlo program computes $N_\pm(n)$, it is better to
convert the asymptotic formula for $p_3(n)$ into  one involving $N_\pm(n)$. We
take the log of Eq. \eqref{ntop} to obtain
\begin{equation}
\log p_3(n) - \log p_3(n-1) = \log N_+(n-1) -\log N_-(n)\ .
\end{equation}
The right hand side of the above equation is something involving $N_\pm(n)$ as
we
wish. We expect that the  asymptotic formula for $p_3(n)$ will be of the form
\begin{equation}
\log p_3(n) \sim \alpha_3 \ n^{3/4}+ \beta_3 \ n^{
 2/4}  + \gamma_3 \ n^{1/4} + \delta_3  \log n + \epsilon_3  \ .
 \label{solidasym}
\end{equation}
The justification of the above form is as follows. From the work of Bhatia et. al.~\cite{Bathia97}, we know that $n^{-3/4}\log p_3(n) \rightarrow \alpha_3$, a constant. Extending a rigorously proved result for plane partitions ($d=2$)~\cite{Cerf01}, we also anticipate that the four-dimensional Ferrers diagram of a random solid partition, at large $n$, will extend the typical distance $\ell\equiv n^{1/4}$ in all four directions symmetrically.\footnote{More generally, for a $d$-dimensional partition, the natural scale is $\ell=n^{1/(d+1)}$.} Thus, we see that $\log p_3(n)$ grows as $\ell^3$. It is natural to interpret the $\ell^3$ term as the first term in a series in $\ell$. For the case of  partitions as well as plane partitions,  a similar structure holds albeit with the vanishing of the next to leading coefficient.  We do not assume the vanishing of the sub-leading terms in Monte Carlo fits for plane partitions as that should be borne out by the data. The log term naturally arises in deriving the asymptotic formula in these cases as well.  Since we carry out fits to the formula for $n\in [50,10100]$ and wish to determine $\alpha_3$ to three decimal places, given that $50^{-3/4}\simeq 0.05$, we see that at low values of $n$, we do need to include  even the constant term. Of course, the accuracy in the determination of other parameters will be clearly less than the one for $\alpha_3$.

%

From our form for the asymptotic formula as given in Eq. \eqref{solidasym},  one can show that for large $n$, one has
\begin{align}
\begin{split}
\log\bigg [\frac{p_3(n)}{p_3(n-1)}\bigg] &= \log\bigg
[\frac{N_+(n-1)}{N_-(n)}\bigg] \\
& \sim \tfrac{3\alpha_3}4\ n^{-1/4}+\tfrac{\beta_3 }2\ n^{-1/2} +\tfrac{\gamma_3 }4\ n^{-3/4} + \delta_3\  n^{-1}\ .
 \end{split}
\label{nasymptotica}
\end{align}
In the above equation, the coefficient of $\alpha_3$ is the leading term in
$\left[n^{3/4}-(n-1)^{3/4}\right]$ for large $n$ and so on. Further,  the constant term $\epsilon_3$
drops out completely and thus the data for $N_\pm(n)$ can estimate four of five
parameters in the asymptotic formula for solid partitions. 
However, we will \textit{not} use the above formula in our fits as we wish later to go back to the formula for $p_3(n)$ to determine the constant $\epsilon_3$. We use
\begin{align}
\begin{split}
\log\bigg
[\frac{N_+(n-1)}{N_-(n)}\bigg] \sim& \ \ 
 \alpha_3 \big[n^{3/4}\big]_3 + \beta_3 \ \big[n^{2/4}\big]_3  
 + \gamma_3 \ \big[n^{1/4}\big]_3 + \delta_3 \big[ \log n\big]_3 \ ,
 \end{split}
\label{nasymptotic}
\end{align}
where by $\big[h(n)]_p$, we mean the first $p$ terms in a power series expansion for $h(n)-h(n-1)$ at large $n$, Thus, one has
\begin{equation}
\big[n^{3/4}\big]_3 = \tfrac34 n^{-1/4} + \tfrac3{32} n^{-5/4} + \tfrac{5}{128}n^{-9/4}\ ,
\end{equation}
and so on. Note that both equations \eqref{nasymptotica} and \eqref{nasymptotic} share the same number of parameters. However, the fit to \eqref{nasymptotic} fixing $(\alpha_3,\beta_3,\gamma_3,\delta_3)$ is the one
that is better suited to estimating $\epsilon_3$ using Eq. \eqref{solidasym} which contributes mostly at small $n$. We also find that a plot of $n^{-3/4}\log p_3(n)$ vs $n$ reveals that the values given by using \eqref{nasymptotic} leads to tiny shift in the fitted curve from the actual data. This is due to the different value of $\epsilon_3$ obtained.  Of course, it does not matter if we only wish to estimate $\alpha_3$ and possibly, the other three constants.

%

\subsection{Results}

We carry out three types of fits\footnote{The fits were carried out using Mathematica's \textit{NonlinearModelFit} using weights given by the statistical error when possible. We have also used \textit{FindFit} for fits without weights.} in order to get estimates of the errors on the parameters.
\begin{enumerate}
\item Fit  data in the range $n\in [50,10100]$ to the four-parameter formula given in Eq. \eqref{nasymptotic}. This determines the values of $(\alpha_3,\beta_3,\gamma_3,\delta_3)$ that we will use along. We quote the result along with statistical errors.
\begin{multline}
(\alpha_3,\beta_3,\gamma_3,\delta_3) = (1.82228\pm 0.00004, 0.06136\pm 0.0008,\\  
0.999\pm 0.008, -0.828\pm 0.003)\ . \label{estimate1}
\end{multline}
The quality of the fit is indicated by the smallness of the statistical errors. These error bars were estimated using the Mathematica fit taking into account the statistical errors estimated for each point.

\item Next we add the term $\big[-\tfrac14 f\ n^{-5/4}\big]$ to the asymptotic formula Eq. \eqref{nasymptotic} and carry out a five-parameter fit  to see how the four parameters 
change.\footnote{This term corresponds to adding  $f\ n^{-1/4}$ to Eq. \eqref{solidasym}.}  We obtain
\begin{multline}
(\alpha_3,\beta_3,\gamma_3,\delta_3,f) = (1.8215\pm 0.0001, 0.088\pm 0.004,\\  
0.60\pm 0.06, -0.51\pm 0.04,1.44\pm 0.19)\ .  \label{estimate1a}
\end{multline}
Notice that the statistical errors have increased slightly. However, the values of $(\alpha_3,\beta_3,\gamma_3,\delta_3)$ have changed a lot more than the statistical error. We use this change in the parameters to set their errors however continuing to use the values obtained in item 1. We thus conclude  that 
\begin{multline}
\boxed{
(\alpha_3,\beta_3,\gamma_3,\delta_3) = (1.822\pm 0.001, 0.06\pm 0.03,
1.0\pm 0.4, -0.8\pm 0.3)\ .}
\end{multline}

\item  By varying the maximum and minimum values of $n$, i.e.,  let $n\in[N_{min},N_{max}]$ for the estimates, we  study the dependence on our estimates for the four parameters in Eq. \eqref{nasymptotic} repeating the fit described in item 1 above. Keeping $N_{min}=50$ but varying $N_{max}$ from $5000$ to $10100$, we find that the changes in the parameters are of the same order as the statistical errors in Eq. \eqref{estimate1}. Next, 
we fit  data in the range $[N_{min},10100]$ to the four-parameter formula given in Eq. \eqref{nasymptotic} by varying $N_{min}$ from $20$ to $100$. The so-obtained values of $\alpha$ are given in Figure~\ref{a:values}, together with its expected value if the Ansatz were correct. We then obtain the average and standard deviation of the 
various parameters, notably $\alpha_3=1.8220 \pm 0.0006$. The errors in the other parameters is also slightly smaller than in item 2 above. 
\begin{figure}[htbp]
\begin{center}
\includegraphics[height=1.6in]{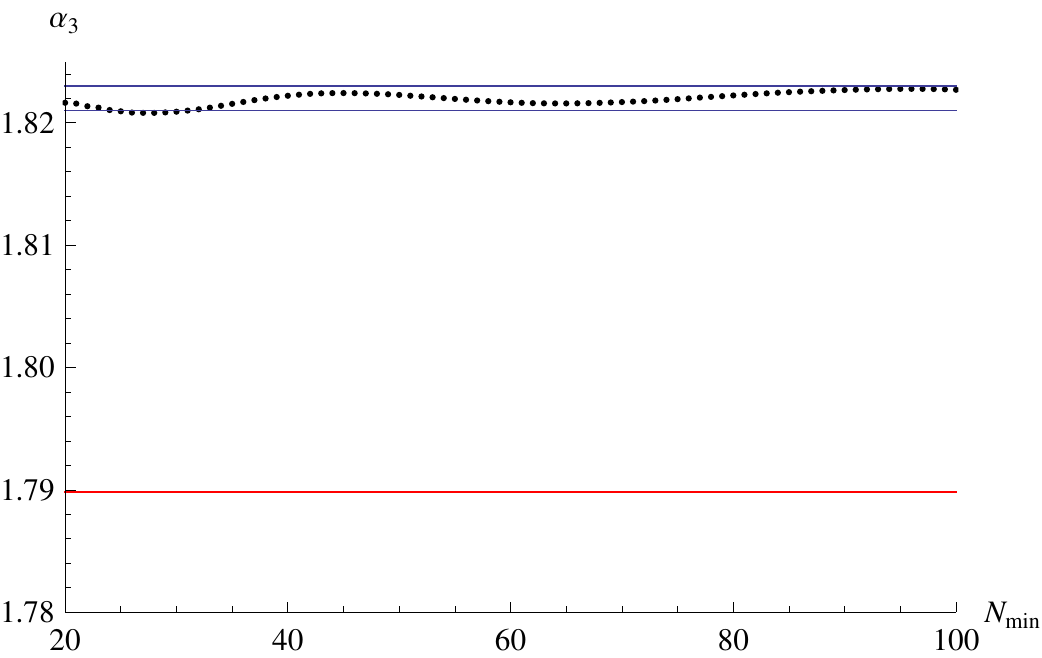}\hfill
\includegraphics[height=1.6in]{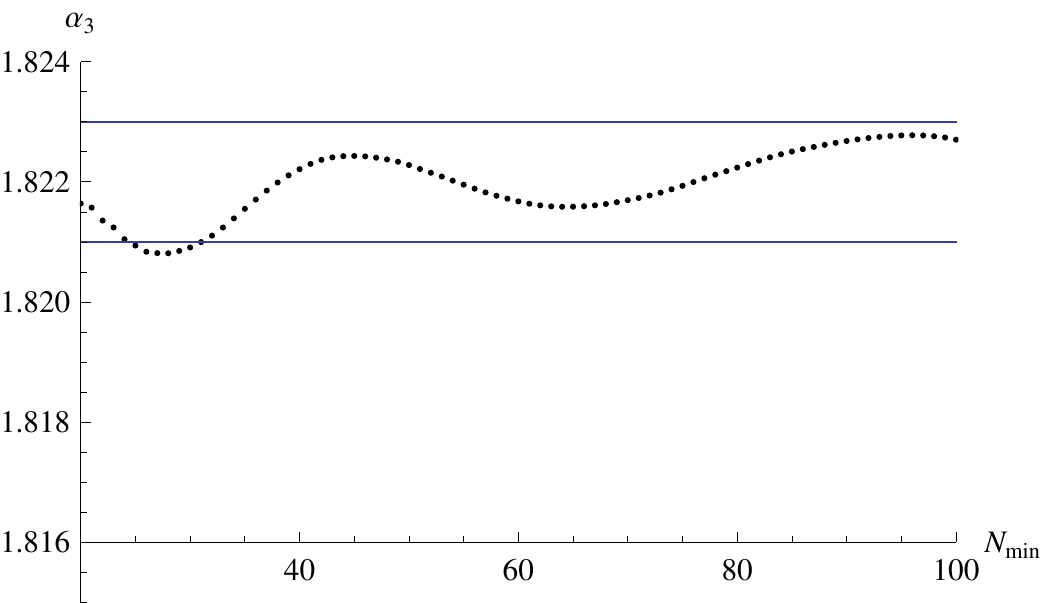}
\end{center}\caption{Fitted values of $\alpha_3$, as defined in Eq.~\eqref{nasymptotic}, in function of $N_{min}$ (dots), together with the conjectured value $1.78982$ if the Ansatz were correct (red horizontal line). The box-size is $B=300$ and $N_{max}=10100$. The second graph to the right is for  the same data but for a different choice of scale for the $y$-axis.  }
\label{a:values}
\end{figure}
The oscillatory character of the fitted values in Figure~\ref{a:values} deserves special attention, because a bad estimate of the actual value of $ \alpha_3$ could be attributed to it, even though it weakens as $N_{min}$ increases. This oscillatory character simply comes from the oscillations of the function $\log p_3(n)$ itself (as displayed in Figure~\ref{delta:values} and discussed below in further detail) which affects the fitted parameter values because the fitting function does not anticipate any subdominant oscillatory term. Their nice close-to-periodic character when plotted in function of $n^{1/4}$, especially at small values of $n$, suggests that these oscillations are not a numerical artifact.

\item We substitute the values of $(\alpha_3,\beta_3,\gamma_3,\delta_3)$ given in Eq. \eqref{estimate1} in Eq. \eqref{solidasym} and then carrying out a one-parameter fit for $n\in[50,100]$ (our best quality data) to obtain $\epsilon_3$.  We obtain $\epsilon_3 = -2.24385$. 
We then add a term $f\ n^{-1}$ to Eq. \eqref{solidasym} and use values given in Eq. \eqref{estimate1a} and carry out a one-parameter fit determine $\epsilon_3$. We obtain $\epsilon_3=-3.12961$. We thus obtain the estimate  
\begin{equation}
\epsilon_3=-2.2\pm0.9\ .
\end{equation}
\end{enumerate}

\subsection{The oscillatory behaviour}

\begin{figure}[htbp]
\begin{center}
\includegraphics[width=2.5in]{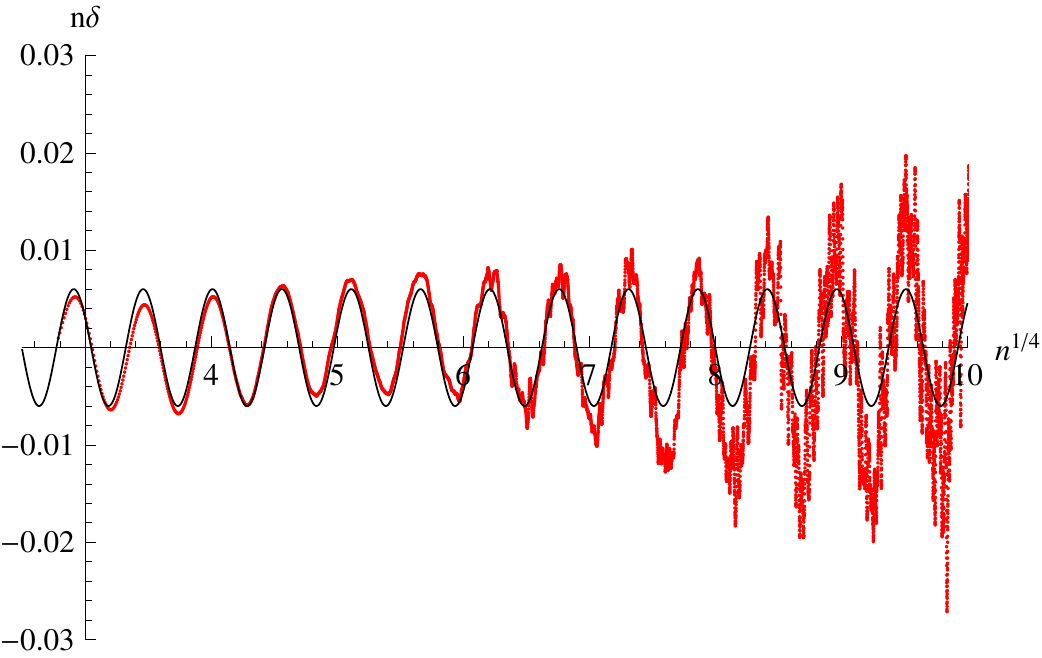}\hfill
\includegraphics[width=2.5in]{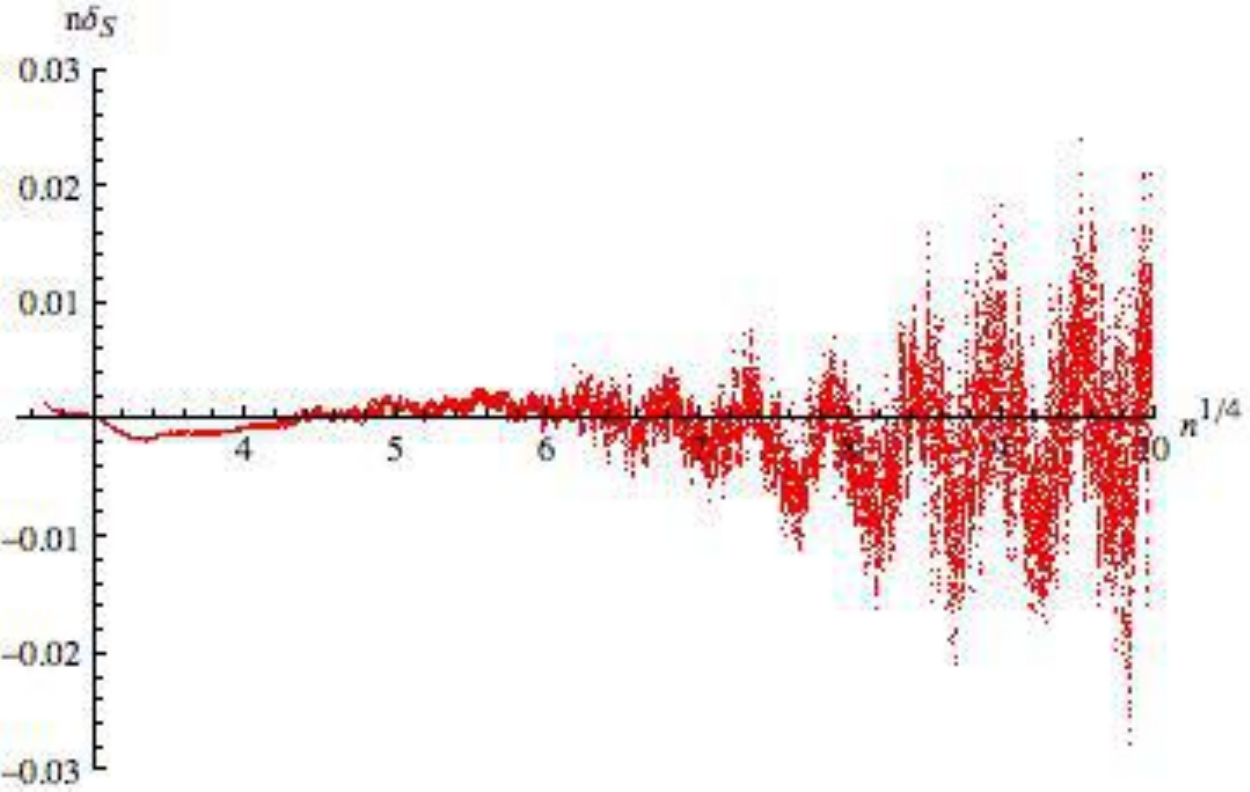}
\end{center}\caption{A plot of the residual $\delta$  multiplied by $n$, as a function of $n^{1/4}$. The box-size is $B=300$,  $N_{min}=50$ and $N_{max}=10000$. The black curve, $0.006 \cos[2\pi(1.817 n^{1/4} - 0.29)]$, correctly reproduces  the periodic character of the numerical data. The graph on the right is the plot of the subtracted residual multiplied by $n$ as a function of $n^{1/4}$.}
\label{delta:values}
\end{figure}

In order to better understand the oscillatory behaviour that we observed in Figure~\ref{a:values}, we define 
\begin{equation}
\delta := \log\big
[\tfrac{N_+(n-1)}{N_-(n)}\big] - \bigg(
 \alpha_3 \big[n^{3/4}\big]_3 + \beta_3 \ \big[n^{2/4}\big]_3  
 + \gamma_3 \ \big[n^{1/4}\big]_3 + \delta_3 \big[ \log n\big]_3\bigg)\ ,
\end{equation}
with $(\alpha_3,\beta_3,\gamma_3,\delta_3)$ taken to be given by Eq. \eqref{estimate1}. The quantity $\delta$ is the \textit{residual} as it measures the difference  between the data and the fit. In the graph to the left in Figure~\ref{delta:values}, we plot $n\delta$ vs $n^{1/4}$. The oscillations are clear to see even when the noise\footnote{The statistical noise in $n\delta$ is given by $n$ times  the estimated statistical error given in Figure~\ref{errors}. Thus it is around $0.13$ for $n=10^4$ but around $0.02$ for $n=6^4$. For $n^{1/4}<4.5$, the magnitude of the oscillations is  larger than the statistical noise. } is larger in magnitude near $n=10^4$.  We find that the oscillatory part can be fitted to the function $ 0.006 \cos [2 \pi (1.817 n^{1/4} - 0.29)]$. We therefore define the \textit{subtracted residual}, $\delta_S$, as follows:
\begin{equation}
 \delta_S :=\ \delta - \tfrac{0.006}{n} \cos [2 \pi (1.817 n^{1/4} - 0.29)]\ .
\end{equation}
The graph to the right in Figure~\ref{delta:values} shows the errors after subtraction. The subtracted residuals are consistent with, albeit somewhat smaller than,  the noise seen in Figure~\ref{errors}. This clearly indicates that the oscillatory behaviour that we observe is not a numerical artifact. 

\begin{figure}[htbp]
\begin{center}
 \includegraphics[width=2.6in]{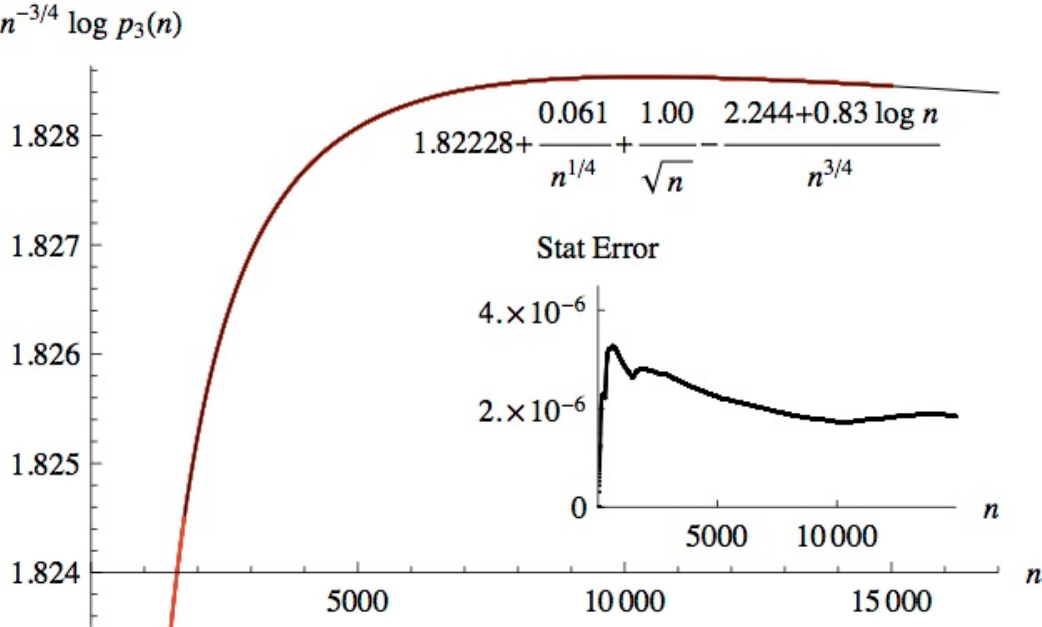}\hfill
\includegraphics[width=2.6in]{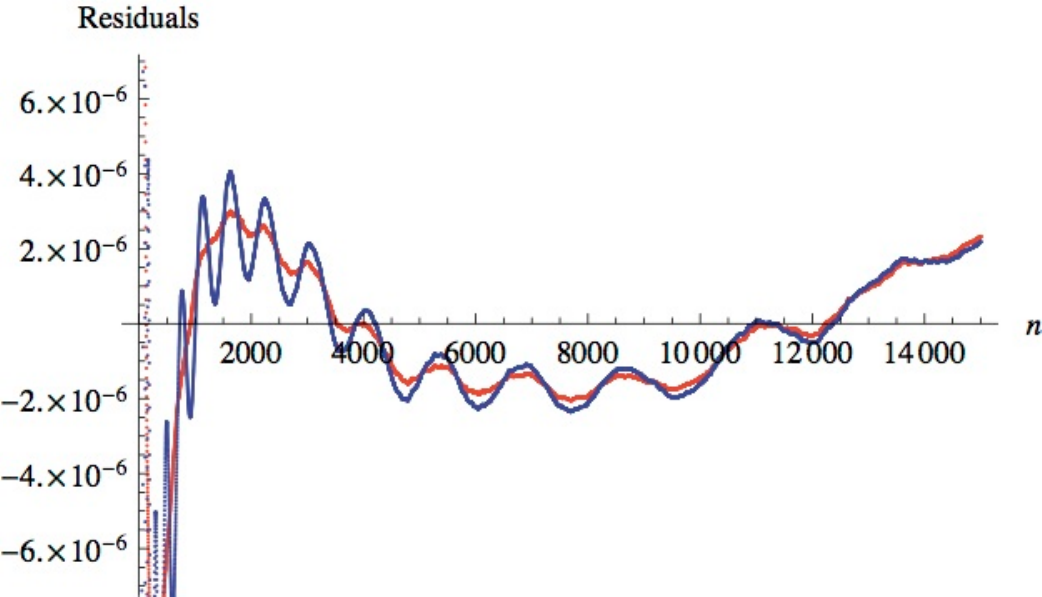}
\end{center}\caption{On the left is the plot of $n^{-3/4}\log p_3(n)$ vs $n$  (as red dots) along with the five-parameter fit (as a black curve). The statistical error in $n^{-3/4}\log p_3(n)$ is shown as an inset.  On the right we plot the residual (in blue) and the subtracted residual (in red) obtained by removing $ n^{-1}\left(-0.0003 + 0.002 \sin[2 \pi (1.817 n^{1/4} -0.29) ]\right)$.} \label{p3twofigs}
\end{figure}

What do  the oscillations  mean for the asymptotic formula Eq. \eqref{solidasym}? It implies that we need to modify it as follows
\begin{multline}
\log p_3(n) \sim \alpha_3 \ n^{3/4}+ \beta_3 \ n^{
 2/4}  + \gamma_3 \ n^{1/4} + \delta_3  \log n + \epsilon_3  \\ 
 + n^{-1/4}\left(f + g \sin[2 \pi \nu n^{1/4} +\varphi ]\right)\ \ .
 \label{solidasym1}
\end{multline}
The additional oscillatory term contributes, to leading order at large $n$,  
$$\frac{\pi g\nu}{2n} \cos [2 \pi \nu n^{1/4} +\varphi ]\, $$ 
to the asymptotic formula given in Eq. \eqref{nasymptotic}. 
Thus we estimate $g=0.0021$, $\nu=1.817$, $\varphi=-0.58\pi$.  The additional non-oscillatory term contributes $\frac{-f}4 n^{-5/4}$ which is of lower order in formula \eqref{nasymptotic}.  In Figure \ref{p3twofigs}, to the left we plot our data for $p_3(n)$ for $n\leq 15000$ (i.e., the full data set) along with our five-parameter fit. On the right, we plot the residuals, one with no subtractions and the second after including the two additional terms given in Eq. \eqref{solidasym1}. The unsubtracted residual shows oscillations while we see  a distinct smoothening in the subtracted residual.

\subsubsection*{Discussion on the possible origin of the oscillatory behaviour}

\begin{figure}[htbp]
\qquad\includegraphics[width=6cm]{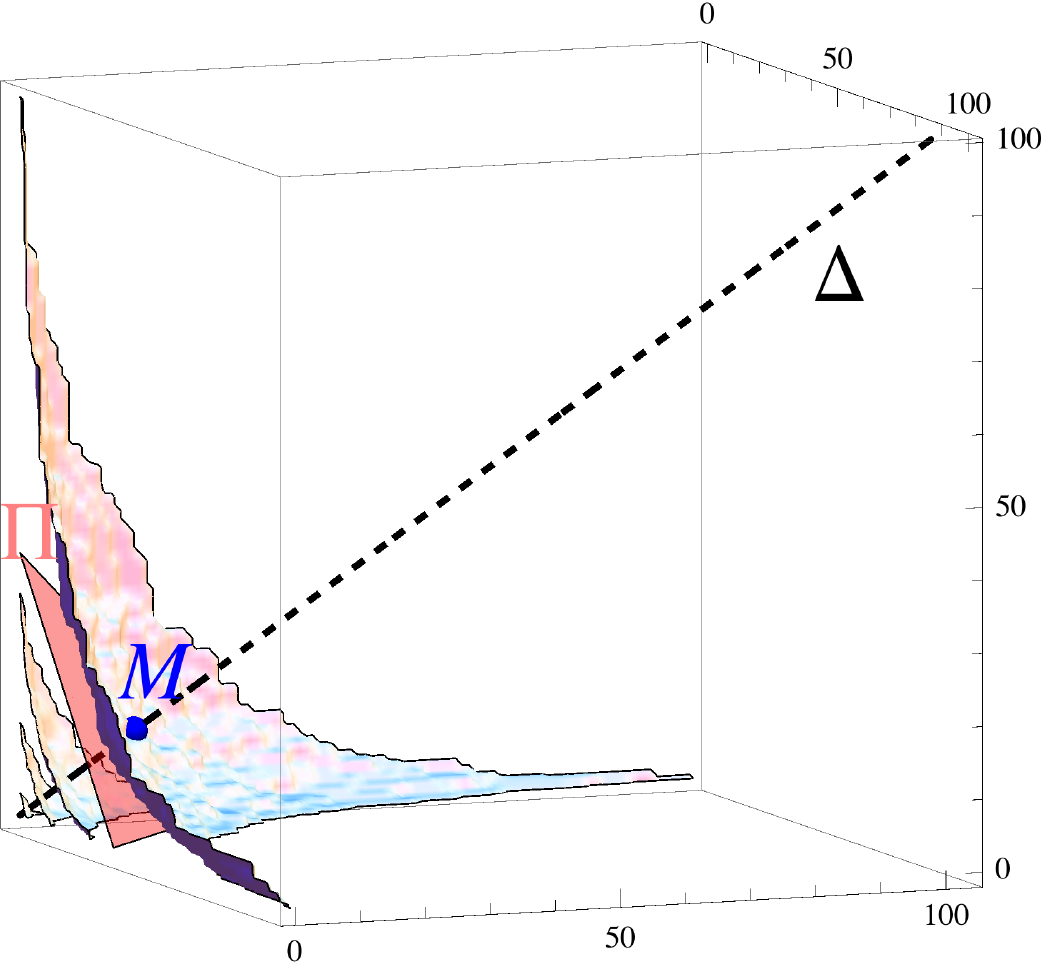}~~~~\includegraphics[width=10cm]{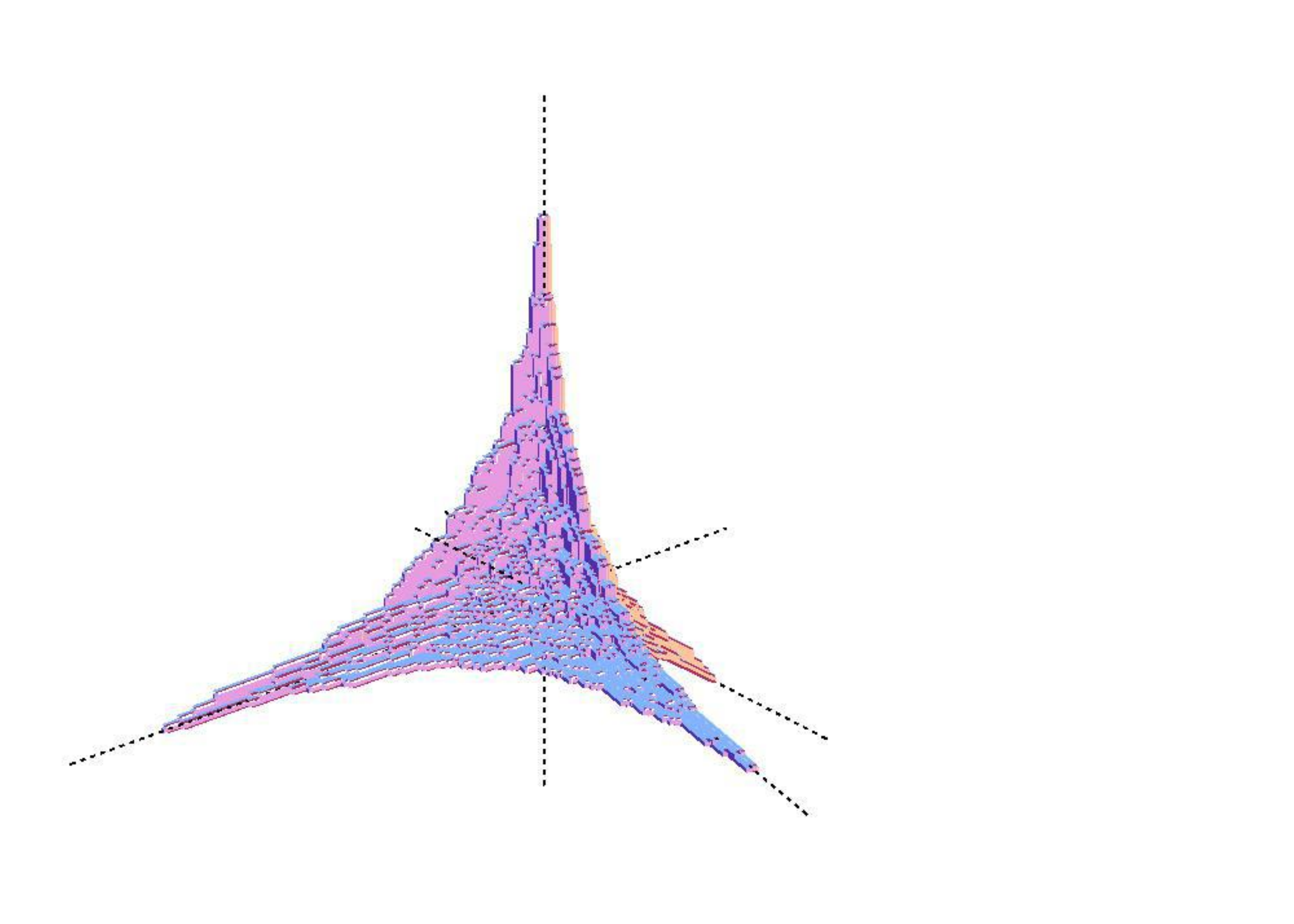} 
\caption{Left: 3D contour-plot representation of a randomly generated solid partition of $n=135000$. As explained in the text [see Eq.~(\ref{parts})], an integer $X_k$ is attached to each vertex of the $\mathbb{N}^{3}$ lattice ($k$ runs over $\mathbb{N}^{3}$); These integers are (weakly) decreasing in each direction of space, and their sum is equal to $n$. The 4 colored curved surfaces separate, from left to right, the 80's from the 81's, the 40's from the 41's, the 20's from the 21's and the 0's from the 1's [i.e. for any $k \in \mathbb{N}^{3}$ on the right (resp. left) of this last surface, $X_k=0$ (resp. $X_k>0$)]. These surfaces are named ``de Bruijn surfaces'' in the tiling context~\cite{Widom02}. The point $M$ (blue dot) belongs to this latest surface and to the line $\Delta$ of equation $x=y=z$ (dashed line). The plane $\Pi$ (in orange) passes through $M$ and is perpendicular to $\Delta$ (see text for more details).
Right: 3D tiling of rhombohedra encoded by the same partition. Only the randomized part of the tiling is represented, the ``frozen'' regions are omitted (see text). The rhombedra edges have unit length and are collinear to the four unit vectors $(\mathbf{e}_i)_{i=1,2,3,4}$ (pointing from the center of a regular tetrahedron to its vertices, not shown). The figure has an overall tetrahedral symmetry, highlighted by the 4 dashed lines collinear to the unit vectors $\mathbf{e}_i$. One of these lines is the image of $\Delta$ in the tiling representation. }
\label{example}
\end{figure}

Without any ambition to definitely solve this issue, which is out of the scope of the present work, we propose that the origin of the oscillations of $\delta$ might be as follows. In Figure~\ref{example} (Left), we have represented a typical partition by using its representation as a collection $(X_k)_{k \in \mathbb{N}^3}$ of integers, as defined in Eq.~(\ref{parts}). We first consider the plane $\Pi$ perpendicular to the $(1,1,1)$ direction passing through $M$ in the figure; by symmetry, this plane is tangent to the average surface separating the 0's and the 1's. We denote by $x_n$ (resp. $y_n$ and $z_n$) the coordinate of its intersection with the $x$ (resp. $y$ and $z$) axis along this axis. By symmetry again, $x_n=y_n=z_n$ and $x_n$ represents the typical linear size of a partition of $n$. It scales like $\ell=n^{1/4}$ (the natural scale of the problem, as discussed above), and we indeed measure numerically that $x_n \simeq 1.88 \; n^{1/4}$. Thus the observation of Figure~\ref{delta:values} (Left) shows that the oscillations are periodic in $x_n$. More precisely, the fitting curve in the figure is of the form $n\delta = \mathrm{const.} \cos[1.817 (2 \pi n^{1/4}+\varphi]$, where $\varphi$ is a dephasing playing no role at this stage, that is to say $n\delta = \mathrm{const.} \cos[0.97(2 \pi x_n)+\varphi']=\mathrm{const.} \cos[2 \pi x_n / \lambda+\varphi']$. The period in $x_n$ is thus $\lambda \simeq 1/0.97\simeq 1$. This strongly suggests that the observed periodicity in fact comes from the periodicity of the underlying $\mathbb{Z}^3$ lattice when the plane $\Pi$ progresses in the $(1,1,1)$ direction as $n$ increases. Thus, depending on the position of the tangent plane $\Pi$ with respect to the lattice vertices, the number of partitions of $n$ seem to deviate weakly from the average value given by Eq.~\eqref{solidasym}. 

This phenomenon can also be discussed in the context of random tilings, which are an alternative representation of solid partitions~\cite{Widom02}. Figure~\ref{example} (Right) provides an example. The figure is the 3D analogous of the ``amoebae'' first explored in the plane partition (or perfect matchings) context~\cite{Cerf01,Kenyon06}. It is obtained by removing from the rhombohedra tiling derived from the left-hand-side partition the ``frozen'' (or periodic) regions where at least one node coordinate is equal either to 0 or to $B$. When $n$ grows, the amoeba expands in a self-similar fashion and the period of the observed oscillations correspond to a displacement of the centers of the 4 amoeba faces along the dashed lines (of $\mathbf{e}_i/3$). This could be related to some putative facetting of amoeba faces near their center. \\
\indent Alternatively, if 3D random tilings are themselves defined through the cut-and-project method~\cite{deBruijn81,Elser85}, this suggests that all cut-hyperplanes perpendicular to the (1, 1, 1, 1) direction in $\mathbb{Z}^4$ do not exactly play the same role, depending on their position relatively to the lattice vertices. In terms of configurational entropy~\cite{Henley91,Destainville98}, there are some preferred positions of the cut-hyperplanes in $\mathbb{Z}^4$. In appendix~\ref{app:bulk:osc}, we indeed observe similar sub-sub-dominant oscillations near the entropy maximum of the boxed solid partition problem (where the amoeba becomes a regular octahedron), which highlights their potentially ubiquitous significance in the solid partition context. This point will have to be confirmed in future studies.

\section{Summary and Concluding Remarks}

We can also address a question first raised in Atkin et. al.\cite{Atkin63}.  \textit{Is $m_3(n)>p_3(n)$ for all $n$? If not, when does it first fail?}  Our data indicates that $p_3(n)$ goes past  $m_3(n)$ near $n= 1425$. We obtain
\begin{equation}
\log m_3(1425)=421.091\quad,\quad \log mc_3(1425)=422.85\pm 0.26\quad.
\end{equation}
In computing $\log mc_3(1425)$, we use formula \eqref{nnaught} with $n_0=72$.
This doesn't depend on the nature of our fits and  appears at values where the statistical errors are small.

Through our transition matrix Monte Carlo simulations,  we saw  that Eq. \eqref{solidasym1} provides a  form for  the asymptotic formula for solid partitions. This can be rewritten as follows:
\begin{multline}
\log p_3(n) \sim \alpha_3 \ \xi^{3/4}+ \beta_3 \ \xi^{
 2/4}  + \gamma_3\ \xi^{1/4} + \delta_3  \log \xi  + \epsilon_3  \\ 
 + \xi^{-1/4}\left(g \sin[2 \pi \nu \xi^{1/4} +\varphi ]\right)+ \cdots \ ,
\end{multline}
where $\xi=n+ \zeta$. The constant $\zeta$ is equivalent to  $f$ in Eq. \eqref{solidasym1} at large $n$.
The new feature is the  appearance of an oscillating term that appears at ``sub-sub-leading order''. In appendix \ref{PPappendix}, we have looked for and not found such oscillations in the plane partition case, that is to say in $\log[p_2(n)/p_2(n-1))]$, which is another manifestation that solid and higher-dimensional partitions display fundamentally different behaviors and cannot be tackled following similar routes (compare for example Refs.~\cite{Widom02,Destainville98}; \cite{Bjorner10}). Beyond our numerical observations, these oscillations will have to be understood. They also potentially represent a guide to future investigations aiming at exactly enumerating solid partitions of an integer because we now know that an enumerating formula will have to contain an oscillating factor. We also note that no oscillations were observed for the MacMahon numbers $m_3(n)$ discussed in appendix \ref{MMasymptotics}.

\begin{figure}[htbp]
\centering
\includegraphics[height=1.7in]{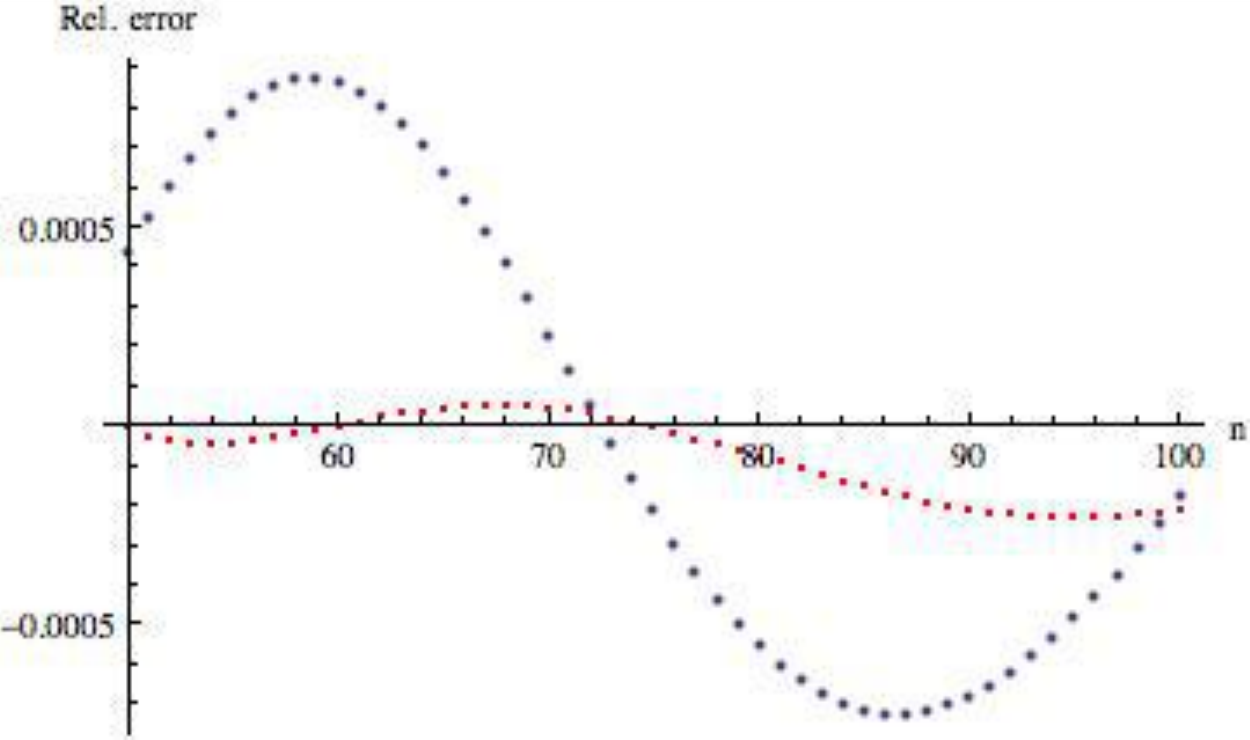}
\caption{The blue points are the relative error (in our estimate for $p_3(n)$)  with  $\zeta=0$ and no oscillatory piece. The red points clearly show the improvement obtained on adding the oscillatory term as well as non-zero $\zeta$.}\label{percentagerror}
\end{figure}

We obtain the following estimates for the various parameters
\begin{equation*}
(\alpha_3,\beta_3,\gamma_3,\delta_3,\epsilon_3) = (1.822\pm 0.001, 0.06\pm 0.03,
1.0\pm 0.4, -0.8\pm 0.3,-2.2\pm0.9)\ .
\end{equation*}
In particular, we see that $\alpha_3$ clearly deviates from the Ansatz value of $1.7898$ thus disproving the conjectures in \cite{Mustonen03,Balakrishnan2011}.  Our value for $\alpha_3$ is within $30\sigma$ of the number quoted in \cite{Mustonen03}. The deviation is around $1.77\%$ which is smaller than the $4.1\%$ deviation observed in the configuration entropy of partitions restricted to a box in \cite{Widom02}. The value of $\beta_3$ is significantly different from the value $0.33$ obtained for the MacMahon numbers $m_3(n)$ (see appendix \ref{MMasymptotics}).  The values for the other parameters are:
\begin{equation}
(\zeta,g,\nu,\varphi)=(-0.000226,1.817,0.00209,-0.268)\ .
\end{equation}
We have not determined the errors in any of them. How good is our asymptotic formula? Since the above parameters really modify the behavior at small $n$, in Figure~\ref{percentagerror}, we plot the percentage error with the exact values of $p_3(n)$ for $n\in [50,72]$ along with Monte Carlo estimates for, i.e., $mc_3(n)$ for $n\in [73,100]$ (this is our best data).  We do it with and without the addition of the oscillatory term and the $\zeta$-shift. Taking these sub-sub-dominant corrections into account clearly improves the fit quality. The unexpected oscillatory term might shed some insight in future works into why partitions in dimensions greater than two do not admit a simple generating function.

Concerning this sub-sub-dominant oscillatory terms, note that if the partitions are represented as Ferrer diagrams in $\mathbb{Z}^4$, we can demonstrate in a similar way as above that the periodicity of $n\delta$ originates from the periodicity of the $\mathbb{Z}^4$ lattice in the $(1,1,1,1)$ direction. If the partitions are now seen as 3D random tilings~\cite{Widom02}, themselves defined through the cut-and-project method \cite{deBruijn81,Elser85}, this original result suggests that all cut-hyperplanes perpendicular to the $(1,1,1,1)$ direction in $\mathbb{Z}^4$ do not exactly play the same role, depending on their position relatively to the lattice vertices. Continuous limit (or coarse-grained) approaches~\cite{Henley91,Destainville98} do not anticipate this fact.

A particularly striking feature of our numerical data, which we have excluded  due to statistical errors, is the occurrence of a maximum of $n^{-3/4}\log p_3(n)$ near $n=10^4$ in Figure~\ref{p3twofigs} (left). Such a break in monotonicity is not exceptional in the context of random tilings, see e.g.~\cite{Widom13}. Having underestimated its role might be at the origin of erroneous conjectures in earlier numerical works.


\bigskip

\noindent \textbf{Acknowledgments:} SG would like to thank Intel India for financially supporting the numerical study of solid partitions.

\appendix

\section{Asymptotics of Plane Partitions}\label{PPappendix}

The asymptotics of plane plane partitions as follows from an application of
Meinardus' formula is\cite{Mutafchiev06,Wright1931}
\begin{equation}
 p_2(n) \sim \frac{(2\zeta(3))^{7/36}}{\sqrt{6\pi}} n^{-25/36} \exp \Big(
\tfrac32 (2\zeta(3))^{1/3} n^{2/3} + \zeta'(-1) \Big) \nonumber
\end{equation}
Thus, one has
\begin{align}
 n^{-2/3}\log p_2(n) &\sim \tfrac32 (2\zeta(3))^{1/3} +
n^{-2/3}\Big(-\tfrac{25}{36} \log n +
\log\tfrac{(2\zeta(3))^{7/36}}{\sqrt{6\pi}} + \zeta'(-1)\Big)\nonumber \\
&\sim 2.00945 -0.694444 n^{-2/3}\log n -1.4631 n^{-2/3}  \label{ppnumbers}
\end{align}
The goal of the Monte Carlo simulation is to reproduce the above formula. In
particular, we should see if we can match the constant term  to one part in $10^3$.

\subsection*{Results from Monte Carlo simulations}

The Monte Carlo simulations for plane partitions are used to estimate the
numbers of plane partitions, say, in the range $[1,N_{max}]$. We then compare these
numbers to the exact numbers and see how the numbers improve with increasing
the statistics by increasing the number of flips. Let $mc_2(n)$ be the values
obtained from the Monte Carlo simulation. We also fit  $n^{-2/3}\log p_2(n)$ 
in the range $[50,2000]$ with the following formula:
\begin{equation}
n^{-2/3} \log p_2(n)\sim \alpha_2\ + \beta_2\ n^{-1/3} + \delta_2\ n^{-2/3} \log n  + \epsilon_2\  n^{-2/3}\ .
\label{ppasymptotic}
\end{equation}
Since our Monte Carlo simulations estimate $N_\pm(n)$, we use
\begin{align}
\begin{split}
\log\big
[\tfrac{N_+(n-1)}{N_-(n)}\big] \sim& \ \ 
 \alpha_2\ \big[n^{2/3}\big]_2 + \beta_2 \ \big[n^{1/3}\big]_2   + \delta_2\ \big[ \log n\big]_2 \ ,
 \end{split}
\label{nppasymptotic}
\end{align}
to extract three of the four parameters $(\alpha_2,\beta_2,\delta_2,\epsilon_2)$ and then determine $\epsilon_2$ using Eq. \eqref{ppasymptotic}.

 \begin{enumerate}
\item Fit  data in the range $n\in [50,2000]$ to the three-parameter formula given in Eq. \eqref{nppasymptotic}. This determines the values of $(\alpha_2,\beta_2,\delta_2)$ that we will use along. We obtain
\begin{equation}
(\alpha_2,\beta_2,\delta_2) = (2.00998, -0.0194366, -0.663683)\ . \label{ppestimate1}
\end{equation}
\item Next we add the term $\big[-\tfrac13 f\ n^{-4/3}\big]$ to the asymptotic formula Eq. \eqref{nppasymptotic} and carry out a four-parameter fit  to see how the three parameters 
change.\footnote{This term corresponds to adding  $f^{-1/3}$ to Eq. \eqref{ppasymptotic}.}  We obtain
\begin{equation}
(\alpha_2,\beta_2,,\delta_2,f) = (2.00923,+0.0120124,-0.731848,-0.41608).  \label{ppestimate1a}
\end{equation}
We take the average of the two sets of numbers and use one half of the difference as an estimate of the error. 
\begin{multline}
(\alpha_2,\beta_2,\delta_2,f) = (2.0096\pm 0.0004, -0.004\pm 0.02,-0.70\pm 0.03,-0.21\pm 0.21)\ .
\end{multline}
 \item We substitute the values of $(\alpha_2,\beta_2,\delta_2)$ given in Eq. \eqref{ppestimate1} in Eq. \eqref{ppasymptotic} and then carrying out a one-parameter fit for $n\in[50,100]$ to determine $\epsilon_3$.  We obtain $\epsilon_2 = -1.44372$. 
We then add a term $f\ n^{-1}$ to Eq. \eqref{ppasymptotic} and use values given in Eq. \eqref{ppestimate1a} and carry out a one-parameter fit determine $\epsilon_2$. We obtain $\epsilon_2=-1.39798$. We thus obtain the estimate  
\begin{equation}
\epsilon_2=-1.423\pm0.025\ .
\end{equation}
 \end{enumerate}
 The  error estimates  for the four parameters through the Monte Carlo simulations  are consistent with the deviation from the exact values.
This  provides some validation for the methods that we used for solid partitions.


\subsection*{No oscillations for plane partitions}

\begin{figure}[htbp]
\begin{center}
 \includegraphics[width=2.6in]{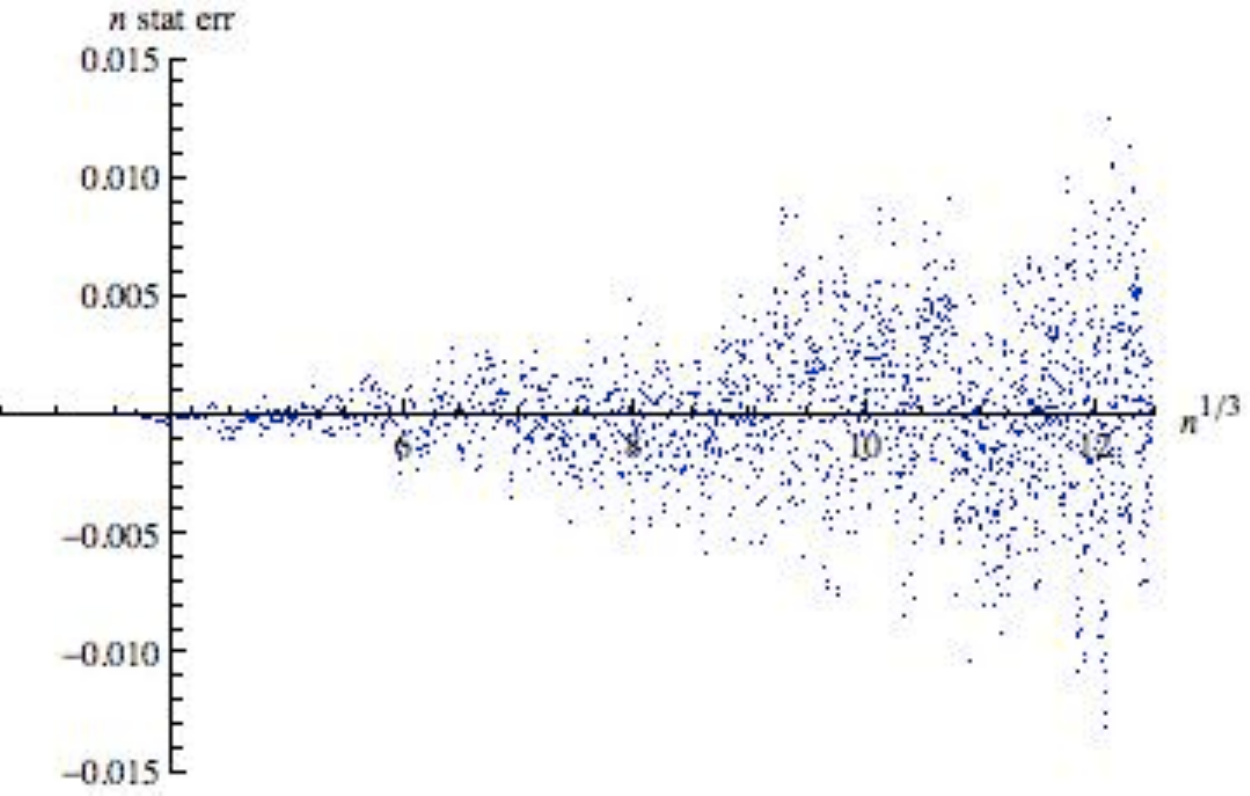}\hfill
\includegraphics[width=2.6in]{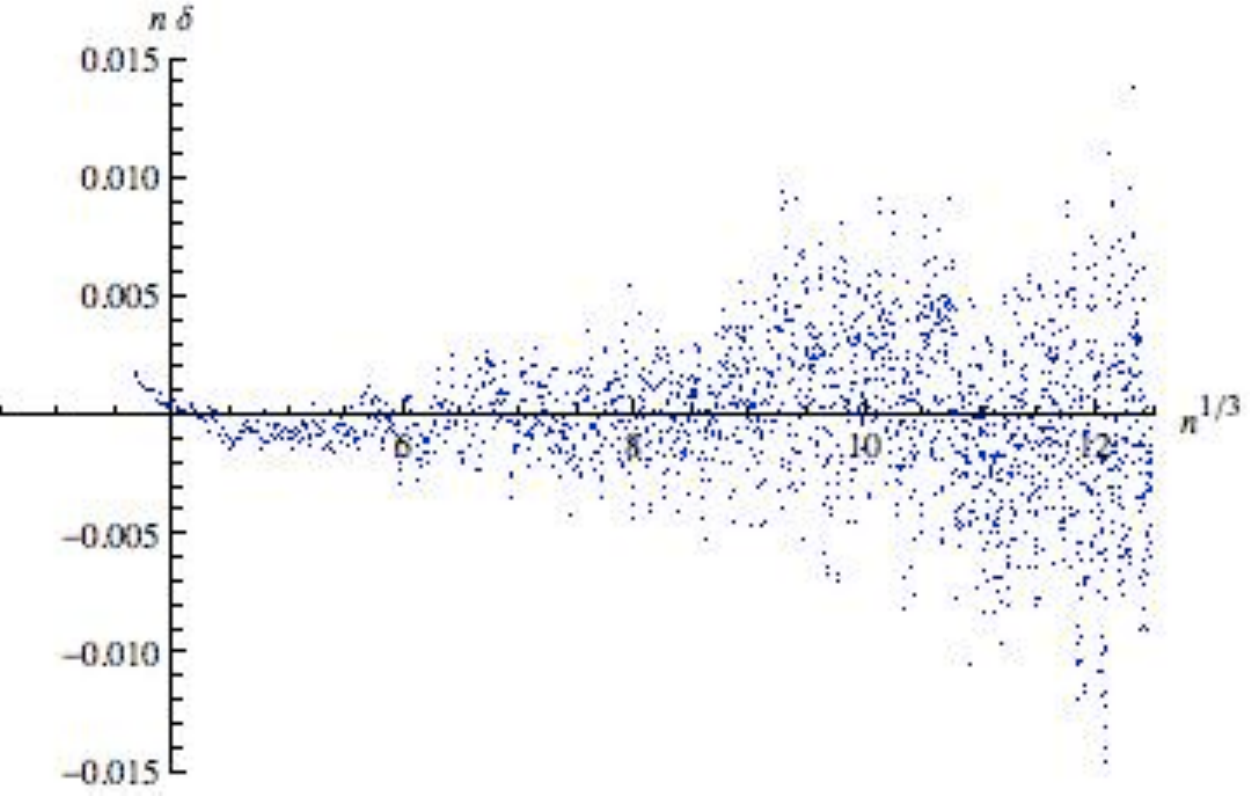}
\end{center}\caption{On the left is the plot of $n$ times the estimated statistical error
vs $n^{1/3}$.  On the right we plot the $n$ times the residual (in blue) to the fit. There are no oscillations to be seen and the residual is comparable to the statistical error.}
\label{PPresidual}
\end{figure}

In order to see if the oscillations that we observe are special to solid partitions, we study the residual in the Monte Carlo data for plane partitions. Since our Monte Carlo simulations did not give us estimates for the statistical errors, we used the difference of  the exact numbers for plane partitions from  the numbers from our Monte Carlo simulations to provide an estimate of the statistical error. In Figure~\ref{PPresidual}, we observe that both the residual and our estimated statistical errors have similar behaviour that is consistent with \textit{no} oscillations.

A similar study on exactly enumerated plane partitions (through Mac Mahon's formula) led us to the same conclusion.



\section{Asymptotics of MacMahon numbers}\label{MMasymptotics}

Recall that the formula that MacMahon guessed for the
generating function of solid partitions gives rise to a series of number that we
call MacMahon numbers, $m_{3}(n)$. One has
\begin{equation}
 {\prod_{n=1}^\infty (1-q^n)^{-\tfrac{n(n+1)}2}}:=\sum_{n=0}^\infty m_3(n)\ q^n\ .\label{MacMahonformula}
\end{equation}
In this section, we briefly discuss the asymptotics of these numbers in order to compare with our results 
for solid partitions.
While these numbers do not have any relation to solid partitions,  we can derive
their asymptotics from the above product formula using Meinardus' method. One
obtains\cite{Balakrishnan2011}
\begin{equation}
n^{-3/4}\log m_3(n) \sim 1.78982 + \frac{0.333546}{n^{
 1/4}}  - \frac{0.0414393}{\sqrt{n}} + \frac{(-1.54436 - 0.635417 \log n)}{n^{3/4}}
\end{equation}

Assuming that their asymptotic behavior is similar to that of solid partitions, 
one can see how well fits to, say  the first 1000 MacMahon numbers,  agree with
the exact asymptotic formula. The accuracy of these fits will provide some hints
towards the quality of the fits that one may expect for  fits using Monte Carlo
data. The fit has five parameters given by the formula
\begin{equation}
n^{-3/4}\log m_3(n) \sim a+ \frac{b}{n^{
 1/4}}  + \frac{c}{\sqrt{n}} + \frac{(e + d \log n)}{n^{3/4}}
\end{equation}
\begin{table}[htbp]
\centering
\begin{tabular}{c|ccccc } \hline 
 & $a$ & $b$ & $c$ & $d$ & $e$ \\ \hline
Fit to $m_3[1000]$ & $1.78909$ & $0.348826$ & $-0.348826$ & $-0.584995$& $-1.42659$\\  \hline
Fit to $m_3[2000]$ & $1.78938$ & $0.343525$  & $-0.142208$ & $-0.598023$   &$-1.4531$ \\ \hline
Exact & $1.78982$ & $0.333546$ & $-0.0414393$ & $-0.635417$ &$-1.54436$ \\ \hline
\end{tabular}
\caption{Results  of the fit to the exact MacMahon numbers $m_3(n)$ in the range
$[50,N_{max}]$.  In column 1, we indicate the value of $N_{max}$ in square
brackets.}
\end{table}
We observe that the the first 2000 numbers reproduce  the leading constant in the asymptotic formula with an error of $0.0005$. Our numerical
study of solid partitions  has used data corresponding to the first 10200 numbers but achieves a slightly lower accuracy. 

\section{Oscillations of boxed solid partitions near the entropy maximum}
\label{app:bulk:osc}

\begin{figure}[htbp]
\begin{center}
\includegraphics[width=8cm]{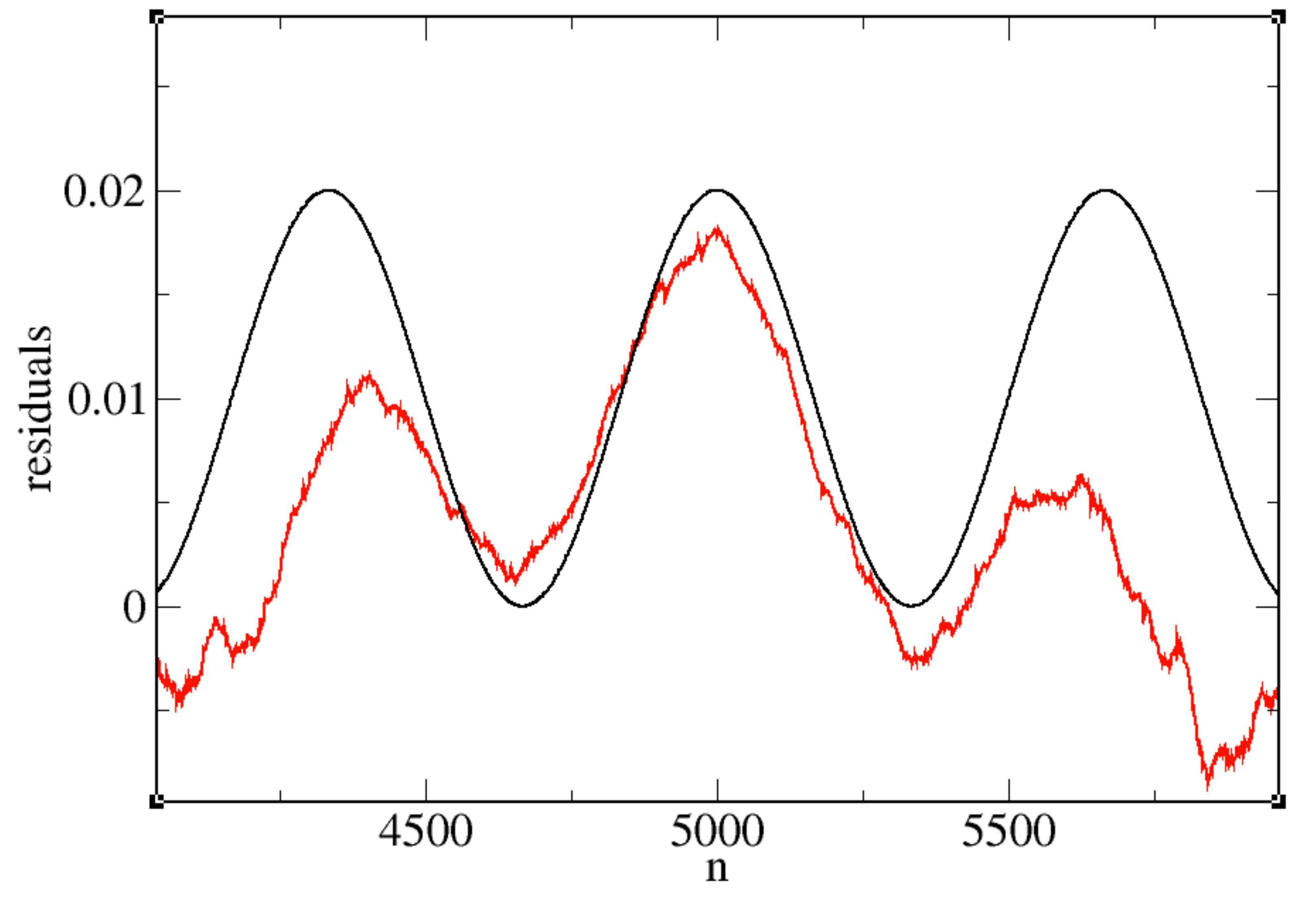}
\end{center}
\caption{Residuals between the numerical entropy $S(n)$ and its fourth-order fit on the interval $[3000,7000]$ for $B=10$; the black sinusoid is a guide for eyes and has period $2 B^3/3\simeq 667$. Numerical data were kindly provided by the authors of Reference~\cite{Widom02}.}
\label{Bulk:oscil}
\end{figure}

So far we essentially focussed on unbounded partitions of an integer, in other words partitions restricted to a box of lateral size $B\gg n^{1/4}$. We needed to have a box for computational reasons, but we have discussed that the box has essentially no incidence. In this context, we have identified original oscillations of the residuals $\delta$, with a period proportional to the natural scale $n^{1/4}$, which we have related to the unexpected sensibility to the underlying lattice. Anticipating that the same phenomenon might have a similar signature for bounded partitions [i.e. partitions restricted to a box of size $B=\mathcal{O}(n^{1/4})$], we have re-examined the numerical data form Reference~\cite{Widom02} as follows. For given values of $B$ and of $n$, there are $p_3(B,n)$ partitions of $n$ such that all coordinates of all nodes are $\leq B$. We then define the partial entropy $S(n) \equiv \log p_3(B,n)$ and we explore its behavior near its maximum (at $n=B^4/2$), by opposition to the limit $n \ll B^4$ (or by symmetry $B^4-n \ll B^4$) studied so far. Near this maximum, it has been conjectured that the amoeba of Figure~\ref{example} (Right) becomes a regular ``arctic'' octahedron~\cite{Linde01,Widom02}. \\
\indent The asymptotic expansion, Eq.~(\ref{solidasym}), is replaced by the fit of $S(n)$ near its maximum by a (somewhat arbitrarily) fourth-order polynomial, and the resulting residual is displayed in Figure~\ref{Bulk:oscil} for $B=10$. Even though it will have to be confirmed in future studies, this figure suggests they there exist oscillations. By analogy with our above findings, we anticipate that there period should be $2B^3/3$. Indeed, as discussed in \cite{Widom02}, near the entropy maximum, the values of $k$ for which the integers $X_k$ (as defined in sections~\ref{background} and \ref{MC}) are generically non-vanishing define a finite subset of $[0,B-1]^3$ (called the ``slab'' in Reference~\cite{Widom02}) of cardinality $2B^3/3$. This ``slab'' is the projection of the ``arctic'' octahedron on $[0,B-1]^3$. Increasing (resp. decreasing) all the integers $X_k$ in the slab by 1 thus leads to an increase (resp. decrease) of $n\equiv \sum_{k\in [0,B-1]^3} X_k$ by $2B^3/3$. The signature of the underlying lattice is thus expected in this case to lead to a sub-sub-dominant correction of period $2B^3/3$ to the entropy $S(n)$, what is indeed suggested by the figure.

\end{document}